\documentclass[fleqn,twoside]{article}
\usepackage{espcrc2}
\usepackage{cite}
\usepackage{amsmath}
\usepackage{eufrak}

\newcommand{\si}{{\rm sign}}

\newcommand{\bq}{\begin{equation}}  
\newcommand{\eq}{\end{equation}}
\newcommand\beq{\begin{equation}}   
\newcommand\eeq{\end{equation}}
\newcommand\bea{\begin{eqnarray}}
\newcommand\eea{\end{eqnarray}}

\newcommand\SH{\,\mbox{$\sqcup \! \sqcup$}\,}

\usepackage{graphicx}
\usepackage[figuresright]{rotating}

\title{Reduction of Multiple Harmonic Sums and Harmonic
Polylogarithms}

\author{J. Bl\"umlein\address{DESY, Deutsches Elektronen
Synchrotron, DESY, Platanenallee 6, D-15738 Zeuthen, Germany}}%
\begin{document}

\begin{abstract}
\noindent
The alternating and non-alternating harmonic sums and other algebraic objects 
of the same equivalence class are connected by algebraic relations which 
are induced by the product of these quantities and which depend on their 
index calss rather than on their value. We show how to find a basis of the 
associated algebra. The length of the basis $l$ is found to be $\leq 1/d$, 
where $d$ is the depth of the sums considered and is given by the 2nd 
{\sc Witt} formula. It can be also determined  counting the {\sc Lyndon} 
words of the respective index set. The relations derived can be used to
simplify results of higher order calculations in QED and QCD.
\vspace{1pc}
\end{abstract}

\maketitle

\section{INTRODUCTION}

\noindent
Multiple nested alternating and non-alternating harmonic 
sums $S_{a_1, \ldots, a_n}(N)$ \cite{HS1,HS2,HS3} 
emerge in perturbative higher order calculations within QED and QCD
for massless fermions,
\begin{eqnarray}
\label{eqHS}
S_{a_1, \ldots, a_n}(N) &=& \sum_{k_1 = 1}^N \sum_{k_2 = 1}^{k_1} \ldots
\sum_{k_n = 1}^{k_{n-1}} \frac{\si(a_1)^{k_1}}{k_1^{|a_1|}}
\nonumber\\ & & \ldots
\frac{\si(a_n)^{k_n}}{k_n^{|a_n|}}~.
\end{eqnarray}
Here, $a_k$ are positive or negative integers and $N$ is a positive
even or odd integer depending on the observable under consideration. One
calls $n$ the {\sf depth} and $\sum_{k = 1}^n
|a_k|$ the {\sf weight} of a harmonic sum. Harmonic sums are associated to
{\sc Mellin} transforms of real functions or {\sc Schwartz}--distributions
$f(x)$~$\epsilon~{\cal S}'[0,1]$~\cite{DISTR}
\begin{eqnarray}
S_{a_1, \ldots, a_n}(N) = \int_0^1 dx~x^N~f_{a_1, \ldots, a_n}(x)
\end{eqnarray}
which emerge in field theoretic calculations. Finite harmonic sums are
related to harmonic polylogarithms $H_{b_1, \ldots, b_n}(x)$~\cite{VR}.
Their $1/(1\pm x)$--weighted {\sc Mellin} transform yields harmonic sums.
The inverse {\sc Mellin} transform relates the harmonic sums to functions 
of {\sc Nielsen} integrals~\cite{NIELS}
of the variable $x$ at least for all sums of weight $w \leq 4$ as shown in      
\cite{HS3}, and associated generalizations for higher weight.
{\sc Nielsen} integrals are a generalization of the usual
polylogarithms~\cite{POLYL}. In the limit $N \rightarrow \infty$ the
convergent multiple harmonic sums, i.e. those where $a_1 \neq 1$, yield
(multiple) Zeta--values $\zeta_{a_1, \ldots, a_n}$, which are also called
{\sc  Euler--Zagier} sums~\cite{EZ}. A generalization of both
harmonic
polylogarithms and the {\sc Euler--Zagier} sums are the nested
$Z$--sums~\cite{MUW}, which form a {\sc Hopf} algebra~\cite{HOPF,KAS}  
and are related to {\sc Goncharov's} multiple
polylogarithms~\cite{GON}.
For a recent review see \cite{WALDS}.

Higher order calculations in massless filed theories are either performed 
in {\sc Mellin}--$N$ space referring to harmonic sums or in the space of 
the momentum fractions $x$ representing the results in terms of {\sc 
Nielsen}--type integrals. The principal complexity is determined by the
amount of possible terms contributing. In the case of the 2--loop 
coefficient functions in momentum fraction space \cite{ZN} 77 different 
functions occurred, cf. 
\cite{HS3}. This number compares to the amount of all possible different  
nested harmonic sums up to weight {$w=4$}, $80 = 3^w-1$. For the 3--loop 
anomalous dimensions~\cite{L3} one expects the contribution of a wide 
class of the
$w=5$ harmonic sums and for the 3--loop coefficient functions of the $w=6$ 
harmonic sums, which means 242 or 728 sums, respectively. These sums are
not independent but connected by different kind of relations. In the 
present paper we summarize a first class of relations recently being 
discussed in Ref.~\cite{ALGEBRA}, the so-called algebraic 
relations.~\footnote{For further relations see~\cite{BLU1}.} It 
turns out that these relations emerge from the index-structure and the 
multiplication relation of the objects considered and are widely 
independent of other properties of the harmonic sums. In this way an 
equivalence class of even more objects is defined having the same 
properties or can be found as special cases thereof. One example is the
set of the harmonic polylogarithms~\cite{VR}. 

To obtain manageable expressions it is of importance to apply all these 
relations through which the number of basic functions to be referred to
is considerably reduced. Experience shows that the {\sc Mellin} space 
representation yields simpler expressions in general~\cite{BM1}, which 
would not be seen easily working in $x$ space. 
To obtain as simple as possible expressions it
is of special importance because of the fact 
that data--analyses require compact results in $x$--space, which can be 
obtained using analytic continuations for the basic sums~\cite{ANCONT} and 
performing a {\sf single} numeric {\sc Mellin} inversion~\cite{BV1} for 
the whole problem. Since the evaluation of precise analytic continuations
needs special effort any possible reduction carried out before is of help.

\section{ALGEBRAIC RELATIONS}

\noindent
The product of two finite harmonic sums (\ref{eqHS}) yields
\small
\begin{eqnarray}
\label{eqPROD}
&&
S_{a_1, \ldots, a_n}(N) \cdot S_{b_1, \ldots, b_m}(N)
\nonumber\\ &&
= \sum_{l_1=1}^N \frac{\si(a_1)^{l_1}}{l_1^{|a_1|}}
  S_{a_2, \ldots, a_n}(l_1)\, S_{b_1, \ldots, b_m}(l_1) \nonumber\\ & &
+ \sum_{l_2=1}^N \frac{\si(b_1)^{l_2}}{l_2^{|b_1|}}
  S_{a_1, \ldots, a_n}(l_2)\, S_{b_2, \ldots, b_m}(l_2) \nonumber\\ & &
- \sum_{l=1}^N \frac{[\si(a_1) \si(b_1)]^l}{l^{|a_1|+|b_1|}}
  S_{a_2, \ldots, a_n}(l)\, S_{b_2, \ldots, b_m}(l)~.
\nonumber\\
\end{eqnarray}
\normalsize
We introduce the {\sf shuffle product} $\SH$ of a single  and a
general finite harmonic
sum
\begin{eqnarray}
\label{eqSTAF}  
&&S_{a_1}(N) \SH S_{b_1, \ldots, b_m}(N)
= S_{a_1, b_1, \ldots, b_m}(N) 
\nonumber\\ & & + S_{b_1, a_1, b_2, \ldots, b_m}(N)
+ \ldots + S_{b_1, b_2,  \ldots, b_m, a_1}(N)
\nonumber\\
\end{eqnarray}
which is a linear combination of the sums of depth $m+1$.
The shuffle product of two harmonic sums
of depth $n$ and $m$, $S_{a_1, \ldots, a_n}(N)$ and $S_{b_1, \ldots,
b_m}(N)$, is  
the sum of all harmonic sums of depth $m+n$ in the index set of which 
$a_i$
occurs left of $a_j$ for $i < j$ and likewise for
$b_k$ and $b_l$ for $k < l$.
As an example the shuffle product of two threefold harmonic sums is given 
by 
\small
\begin{eqnarray}
\label{eqD6C}
& &S_{a_1, a_2, a_3}(N) \SH S_{a_4, a_5, a_6}(N) = \nonumber\\ 
& &
S_{a_1, a_2, a_3, a_4, a_5, a_6}(N)
     +S_{a_1, a_2, a_4, a_3, a_5, a_6}(N) \nonumber\\ & & 
     +S_{a_1, a_2, a_4, a_5, a_3, a_6}(N)   
     + S_{a_1, a_2, a_4, a_5, a_6, a_3}(N) \nonumber\\ & &  
     +S_{a_1, a_4, a_2, a_3, a_5, a_6}(N)  
     +S_{a_1, a_4, a_2, a_5, a_3, a_6}(N) \nonumber\\  & &
     + S_{a_1, a_4, a_2, a_5, a_6, a_3}(N)  
     +S_{a_1, a_4, a_5, a_6, a_2, a_3}(N) \nonumber \\ & & 
     +S_{a_1, a_4, a_5, a_2, a_6, a_3}(N)   
     + S_{a_1, a_4, a_5, a_2, a_3, a_6}(N) \nonumber\\ & &
     +S_{a_4, a_5, a_6, a_1, a_2, a_3}(N)
     +S_{a_4, a_5, a_1, a_6, a_2, a_3}(N) \nonumber\\ &&
     + S_{a_4, a_5, a_1, a_2, a_6, a_3}(N)
     +S_{a_4, a_5, a_1, a_2, a_3, a_6}(N) \nonumber\\ & &
     +S_{a_4, a_1, a_5, a_6, a_2, a_3}(N) 
     + S_{a_4, a_1, a_5, a_2, a_6, a_3}(N) \nonumber\\ & &
     +S_{a_4, a_1, a_5, a_2, a_3, a_6}(N)
     +S_{a_4, a_1, a_2, a_3, a_5, a_6}(N) \nonumber\\ &&
     + S_{a_4, a_1, a_2, a_5, a_3, a_6}(N)
     +S_{a_4, a_1, a_2, a_5, a_6, a_3}(N) 
\nonumber\\
\end{eqnarray}
\normalsize
Finally one establishes a system of linear equations in which the 
linear elements of the shuffle products form the variables and a 
polynomial out of harmonic sums of lower depth forms the inhomogeneity. We 
furthermore consider all index permutations. This system contains all 
algebraic relations. In Ref.~\cite{ALGEBRA} all solutions for harmonic 
sums up to depth $d=6$ were given. This complies to the level of 
sophistication needed to reduce the corresponding relations which emerge 
for massless 3--loop coefficient functions.

\section{NUMBER OF ALGEBRAICALLY INDEPENDENT HARMONIC SUMS}

\noindent
Let us consider the index set of a harmonic sum of depth $d$. One may 
consider this set as a {\sf word} $w$ or a non--commutative product of 
{\sf 
letters} of an {\sf ordered alphabet}  ${\mathfrak A} = \{a,b,c,d, 
\ldots\}$. Any word can be decomposed into three parts
\begin{equation}
w = pxs~,
\end{equation}
a prefix $p$, a suffix $s$, and the remainder part $x$. Among all words 
$w$ 
the {\sc Lyndon} words, cf. e.g.~\cite{REUT}, are those being smaller than 
any of its suffixes.

According to a Theorem by {\sc Radford}~\cite{RADF} the shuffle algebra 
discussed above is freely generated by the {\sc Lyndon} words, i.e. the 
length of its basis is given by the number of {\sc Lyndon} words. We would 
like to count the number of {\sc Lyndon} words for index sets, where 
the same letters can emerge repeatedly. The corresponding relation is due
to {\sc Witt}~\cite{WITT} and will be called  2nd {\sc Witt} formula
\begin{eqnarray}
\label{eqWIT2}
l_n(n_1, \ldots, n_q) &=& \frac{1}{n} \sum_{d|n_i} \mu(d)
\frac{\left(\frac{n}{d}\right)!}
{\left(\frac{n_1}{d}\right)! \ldots \left(\frac{n_q}{d}\right)!},
\nonumber\\
\end{eqnarray}
with $n = \sum_{k=1}^q n_k$.
Here $\mu(d)$ denotes the {\sc M\"obius} function.
One may derive $l_n(n_1, \ldots, l_q)$ using the generating functional 
\begin{eqnarray}
\label{eqWIT2GF}
\frac{1}{1-x_1 - \ldots - x_{n_q}} =  \prod_{n = 1}^\infty
\left(\frac{1}{1-\sum_{k=1}^q x_k^{d_k}}\right)^{l_n(n_i)}
\nonumber
\end{eqnarray}
Note that (\ref{eqWIT2}) is related to the {\sc Gauss-Witt} relation 
mentioned by {\sc Hoffman}~\cite{HOF} for the number of basic multiple 
Zeta--values of weight $w$ for $\forall n_i > 0$ if all cases for fixed 
weight are summed over. An even more strict 
relation in the inclusive case has been conjectured by {\sc 
Zagier}~\cite{EZ} and {\sc 
Broadhurst} and {\sc Kreimer}~\cite{BRKR} in the case of multiple 
Zeta--values and verified up to $w=12$.

Let us come back to Eq.~(\ref{eqWIT2}). We can draw some immediate 
conclusions out of this relation. If the numbers $n_i$ have no common 
divisor larger than 1, the number of the basis elements compared to the 
number of all objects equals $1/d$, where $d$ denotes the {\sf depth} 
of the index set. In case of common divisors larger than 1 we checked that 
the basis is always shorter for all depths up to $d=10$, see 
\cite{ALGEBRA}. 

\begin{center}  
\begin{tabular}[h]{||r||r|r|l||}
\hline \hline %
\multicolumn{1}{||c||}{\sf Weight}&
\multicolumn{1}{c|}{\sf  \# Sums}&
\multicolumn{1}{c|}{\sf  \# Basic Sums}&
\multicolumn{1}{c||}{\sf Fraction  }\\
\hline\hline  
1 &   2 &   0 &    0.0    \\
2 &   8 &   1 & 0.1250 \\
3 &  26 &   7 & 0.2692 \\
4 &  80 &  23 & 0.2875 \\
5 & 242 &  69 & 0.2851 \\
6 & 728 & 183 & 0.2513 \\
\hline \hline
\end{tabular}
\renewcommand{\arraystretch}{1.0}
\end{center}

\section{CONCLUSIONS}

\noindent
The product of finite alternating or non--alternating harmonic sums is given
by
the shuffle product of harmonic sums and polynomials of harmonic sums of 
lower
depth. These representations imply algebraic relations between the harmonic
sums. If one considers all harmonic sums associated to an index set
$\{a_1, \ldots, a_k\}$ one may express these sums by a number of basic
sums. It turns out that this number is given by the 2nd {\sc Witt}
formula which counts the
number of {\sc Lyndon} words corresponding to the respective index set.
The set of these {\sc Lyndon} words generates in this sense all harmonic
sums of this class {\sf freely}. By solving the corresponding linear
equations we
derived the explicit representation of all harmonic sums up to depth $d=6$
without specifying the indices numerically and gave all expression which
are structurally needed to express the sums up to weight $w=6$. The
counting relations for the basis of the finite harmonic sums were given
up to depth $d=10$. The relations derived hold likewise for other
mathematical objects obeying the same multiplication relation or a simpler
one, which is being contained, as that for harmonic polylogarithms. This
is due to the fact that the relations derived depend on the index set
and the multiplication relation but on no further properties of the
objects considered.

The ratio of the number of basic sums for a given index set and
the number of all sums is mainly determined by the depth $d$ rather
than the weight of the respective sums, due to the pre-factor $1/d$ in the
{\sc Witt} formula. Modifications occur due to common non-trivial divisors
of the numbers of individual indices in the set being considered. Up to
$d=10$ we showed that the fraction of basic sums is always $\leq 1/d$
compared to all sums. The use of these algebraic relations leads to a
considerable reduction in the set of functions needed to express the
results of higher order calculations in massless QED and QCD and related
subjects. For practical applications such as the description of the QCD
scaling violation of the structure functions in deeply inelastic
scattering the harmonic sums  occurring in the {\sc Mellin} $N$ space
calculation have to be translated to $x$--space by the inverse {\sc
Mellin} transform. For this reason the respective harmonic sums have to be
analytically continued in the argument $N$ to complex values, which
requires a high effort using numerical procedures. It is therefore
recommended to use as many as possible relations between the $N$ space
objects before to perform the last step only for the reduced set.

\vspace*{1mm}
\noindent
{\bf Acknowledgment.}~~
This paper was supported in part by
DFG Sonderforschungsbereich Transregio~9, Computergest\"utzte Theoretische
Physik.


\end{document}